\documentclass[a4paper,11pt,notoc]{JHEP3}

\usepackage{
graphicx
,epsfig
,ifthen
,amsmath
}
\graphicspath{{./}}

\newcommand{\beq}{\begin{equation}}
\newcommand{\eeq}{\end{equation}}
\newcommand{\beqn}{\begin{eqnarray}}
\newcommand{\eeqn}{\end{eqnarray}}

\newcommand{\nn}{\nonumber}

\newcommand{\ov}{\overline}

\newcommand{\be}{\beta}

\newcommand{\eps}{\epsilon}

\newcommand{\la}{\lambda}

\newcommand{\eg}{{\em e.g.} }
\newcommand{\ie}{{\em i.e.} }

\newcommand{\D}{\Delta}

\def\simge{\mathrel{\rlap{\raise 0.511ex \hbox{$>$}}{\lower 0.511ex \hbox{$\sim$}}}}
\def\simle{\mathrel{\rlap{\raise 0.511ex \hbox{$<$}}{\lower 0.511ex \hbox{$\sim$}}}} 

\title{\boldmath Large $|V_{ub}|$: A challenge for \\ the Minimal Flavour Violating MSSM}

\author{Wolfgang~Altmannshofer, Andrzej~J.~Buras,
Diego~Guadagnoli and Michael~Wick\\ 
Physik-Department, Technische Universit\"at M\"unchen,\\
D-85748 Garching, Germany}

\abstract{
\noindent Under the assumption of Minimal Flavour Violation (MFV), the Unitarity Triangle
(UT) can be determined by using only angle measurements and tree-level observables. In this respect,
the most accurate quantities today available are $\sin 2 \beta_{\psi K_S}$, $|V_{cb}|$
and $|V_{ub}|$. Among the latter, $|V_{ub}|$ is at present the quantity suffering from largest
systematic uncertainties, given the discrepancy between the inclusive and the exclusive
determinations. 

\noindent We show with a numerical fit how sensitively the MFV-UT determination depends
on the choice of $|V_{ub}|$. In addition, we focus on the implications of the inclusive
value for $|V_{ub}|$, which favors two non-SM like solutions in the $\ov \rho - \ov \eta$
plane. We study in detail the possibility of reproducing such solutions within the MFV MSSM. 
Our findings indicate that the case for the MFV MSSM is in this respect quite problematic,
unless the non-perturbative parameters $\xi$ and $B_K$ are significantly different
from those obtained by lattice methods. As a byproduct, we point out that
scenarios with 200 GeV $\lesssim M_A \lesssim$ 500 GeV and $\tan \beta \simeq 50$ that
predict a significant suppression for $\D M_s$ in {\em correlation} with an enhancement 
for BR$(B_s \to \mu^+ \mu^-)$ have to be fine-tuned in order not to violate the new combined 
bound on the latter decay mode from the CDF and D{\O} collaborations. Relatively large
correlated effects can however still occur for negative values of $\mu$ and large values 
for $M_A \gtrsim 500$ GeV, increasing with increasing $\tan \beta \gtrsim 30$.
}
\preprint{TUM-HEP-670/07}

\begin{document}

\section{Introduction} \label{sec:intro}

Most of the calculable models of low-energy New Physics (NP) available at present 
tend to be invasive in the sector of flavour violation and to destroy the very specific 
pattern of flavour changing neutral current (FCNC) effects predicted by the SM. In the example
of SUSY, this happens because of its soft sector. In fact, in absence of compelling
symmetries restricting the form of the soft terms, the latter must be parameterized most 
generally. However, when calculating SUSY contributions to flavour observables with such 
generic soft terms, it is hard to hide simultaneously all their effects behind the 
already small SM predictions, which agree rather well with the experimental data. The soft terms 
parameter space that turns out to survive FCNC constraints looks then very fine tuned and this 
raises the SUSY flavour problem.

A different, phenomenologically coherent approach, is to assume -- on the basis of the
success of the SM description of FCNC processes -- that the flavour sector of any NP
model maintains a natural mechanism of ``near-flavour-conservation''
\cite{hall-randall-MFV}. The latter would then allow small effects in exact analogy with what 
happens for the SM. 

This idea has been first elucidated as a meaningful requirement for models of NP at the
EW scale by the authors of Refs. \cite{chivukula-georgi-MFV,hall-randall-MFV}. It has then
been formalized in \cite{MFV} as a consistent effective field theory framework, called
Minimal Flavour Violation (MFV). In this framework, the SM mechanism of near-flavour-conservation 
is extended to NP by means of the SM Yukawa couplings. Since the latter are in the SM the {\em only} 
sources of flavour breaking, in MFV they play the role of fundamental ``building blocks'' of flavour 
violation and every new source of flavour violation, entering the NP Lagrangian, becomes function of 
them.

This approach is more general than the so-called constrained MFV (CMFV)
\cite{BurasMFV,BurasZakopane,BBGT}, in which one also imposes the same operator
structure as in the SM.

Recently, the MFV framework of \cite{MFV} has been applied to the MSSM at low 
$\tan \beta$, in a detailed numerical study of meson mixings \cite{ABG}.\footnote{The same approach, 
though in a context where effects beyond CMFV are not visible, has been adopted in Ref.
\cite{IMPST}, where the decays $K_L \to \pi^0 \nu \ov \nu$, $K^+ \to \pi^+ \nu \ov
\nu$ and $K_L \to \pi^0 \ell^+ \ell^-$ have been studied.}
Thereby, the phenomenological differences between MFV and CMFV have been spelled 
out. It has also been shown how the implementation of the MFV limit makes SUSY contributions 
to meson mixings {\em naturally} small, even for a SUSY scale of a few hundreds GeV.

If NP is of MFV nature -- as the lesson of $B$-factories seems to hint -- the
findings of Ref. \cite{ABG} indicate that the search for NP effects in flavour 
observables is more challenging, but not less important, since MFV models become more
predictive. In MFV, the focus is on small, but in many cases visible effects. With increased precision 
of theoretical methods and the amount of data soon available from the LHCb and later,
hopefully, from the Super-B \cite{SuperB}, one will have such a level of cross-check among different channels that 
NP effects of even MFV nature should become visible.

An important test of overall consistency among different flavour processes is certainly the 
global fit to the CKM matrix, which assumes the SM in all observables. However, it is also important 
to consider fits where one restricts to specific classes of observables (in particular, suitable ratios
thereof) that, when assuming a given NP framework besides the SM, are affected in a controlled way 
or are not affected at all. A first example of this is precisely CMFV \cite{BurasMFV,BurasZakopane,BBGT}, 
that has been analyzed in detail by the UTfit collaboration \cite{UTfit}.

However, as first pointed out in \cite{ABG}, the UT analysis suggested by \cite{BurasMFV}
does not account for the most general framework of MFV. Ref. \cite{ABG} showed in
fact that the definition \cite{MFV} of MFV does not preclude the ratio $\D M_d/\D M_s$ --
included in the CMFV UT analyses -- to be different from the SM value, and therefore, that
it should not be included in MFV UT fits.

As a first aim of this paper, we carry out a MFV fit to the UT. When assuming MFV, the only 
quantities allowed to enter the fit are angle determinations and measurements of tree-level 
processes. Concerning the former, we restrict to $\sin 2 \beta$ as measured from $B
\to \psi K_S$, while we do not include $\alpha$ and $\gamma$, for reasons to be explained
below. As for tree-level processes, we instead restrict to the semileptonic $B$ decays allowing 
to access $|V_{ub}|$ and $|V_{cb}|$. In particular, while the value of $|V_{cb}|$ is quite well established, 
the same cannot yet be said about $|V_{ub}|$, whose inclusive and exclusive determinations are in some 
disagreement with each other, possibly signaling the presence of an underestimated systematic error in
either of the two determinations.

It was noted in ref. \cite{UTfit-PRL} that, if one uses the inclusive value for $|V_{ub}|$ in 
a global UT analysis and parameterizes the presence of NP in a model independent way \cite{UTfit-NP}, 
the fit adjusts the tension between $\sin 2 \beta$ and the `too high' value for $|V_{ub}|$ by 
introducing a small negative new phase $\phi_{B_d} = -2.9^\circ \pm 2.0^\circ$ in the
$B_d$-mixing (see also \cite{BBGT,BFRS-PRL}).

In the present work, we adopt a somehow complementary point of view, since we focus on MFV. 
The definition of MFV \cite{MFV} precludes the existence of new CP violating phases beyond the
CKM one.
In this framework, we discuss the impact of both the exclusive and the inclusive averages for
$|V_{ub}|$ on the UT determination. We show in particular how the inclusive $|V_{ub}|$ favors 
two non-SM like solutions in the $\ov \rho - \ov \eta$ plane.

Pursuing the possibility that the inclusive $|V_{ub}|$ determination be the correct one, 
the next question is whether one can find a MFV extension of the SM able to produce either 
of the two non-SM solutions mentioned above. In particular, one can consider the shifts in 
the value of the side $R_t$ implied by the two solutions. Since $R_t$ is (in the SM) related to 
the ratio $\D M_d / \D M_s$ between the $B_d$- and the $B_s$-system oscillation frequencies, 
one can translate the above shifts into required values for the NP contributions to the ratio itself.
We consider the explicit example of the MFV MSSM. Our findings indicate that the model 
is able to produce the required amount of corrections only in certain fine-tuned regions
of the parameter space. This conclusion holds, barring a substantial shift (above 2 standard 
deviations) in the present central values for the low-energy parameters $\xi$ and $B_K$.

\section{MFV fit of the UT} \label{sec:MFV-UT}

We now turn to our first task, \ie the determination of the UT when MFV is assumed. 
In particular we will focus on the UT side $R_t$, which will be the relevant quantity for our 
subsequent analysis. We have today a whole host of observables, which bear dependence 
on certain combinations of the CKM matrix entries. Hence the determination of the UT apex 
$(\ov \rho, \ov \eta)$ -- and all related quantities -- follows from a 
global fit \cite{CKMfitter, UTfit}, which can include all or a subset of such observables.

We consider three fits of the UT, namely
\begin{itemize}
\item[\bf 1.] a SM fit, including all the `classical' constraints,
\item[\bf 2.] a MFV fit I, including only tree-level observables and 
$\sin 2 \beta_{\psi K_S}$,
\item[\bf 3.] a MFV fit II, analogous to the MFV fit I, but keeping only the inclusive averages
for both $|V_{ub}|$ and $|V_{cb}|$.
\end{itemize}

Concerning the full SM fit, it can be performed by combining all the available
experimental information (see \cite{CKMfitter, UTfit}) and assuming the SM. 
The state-of-the-art results for the SM determination of $R_t$ from the CKMfitter and 
UTfit collaborations (95\% CL) read
\beqn
\label{SM-Rt}
\mbox{CKMfitter:  }(R_t)_{\rm SM} = 0.868 ^{+0.118}_{-0.049}~,
~~~\mbox{UTfit:  }(R_t)_{\rm SM} = 0.906 \pm 0.062~,
\eeqn
in very good agreement with each other. We have performed our SM global fit,
using the {\tt CKMfitter} package \cite{CKMfitter,CKMfitter-paper}, a publicly available FORTRAN 
framework allowing CKM analyses in various statistical approaches, \eg the frequentist one, 
used in the present work. We took advantage of the {\tt CKMfitter} package also for the other
fits performed in the present work. It would be interesting to perform an analogous analysis 
using the UTfit code, which is however not yet publicly available. The input parameters
and constraints used in our global fit is listed in Tables \ref{tab:theo-input} and
\ref{tab:UT-constraints}. For $R_t$ we find (95\% CL)
\beqn
\label{SM-Rt-reference}
(R_t)_{\rm SM} = 0.892 ^{+0.112}_{-0.068}~,
\eeqn
in good agreement with the findings in eq. (\ref{SM-Rt}). Eq.
(\ref{SM-Rt-reference}) will be taken as reference figure in our subsequent analysis.
\TABLE[!t]{
\begin{tabular}{clcl}
\hline
Parameter & \hspace{8ex}Value & Parameter & \hspace{8ex}Value \\
\hline
$\overline{m}_c(m_c)$ & $(1.24 \pm 0.037 \pm 0.095)$ GeV & $G_F$ & $1.16639 \times 10^{-5}$ GeV$^{-2}$ \\
$\overline{m}_t(m_t)$ & $(162.3 \pm 2.2)$ GeV & $f_K$ & $(159.8 \pm 1.5)$ MeV \\
$m_{K^+}$ & $(493.677 \pm 0.016)$ MeV & $B_K$ & $0.79 \pm 0.04 \pm 0.09$ \\
$\D m_{K}$ & $(3.4833 \pm 0.0066)\times 10^{-12}$ MeV & $\alpha_s(m_Z^2)$ & $0.1176 \pm 0.0020_{\rm unif}$ \\
$m_{B_d}$ & $(5.2794 \pm 0.0005)$ GeV & $f_{B_d}$ & $(0.191 \pm 0.027)$ GeV \\
$m_{B_s}$ & $(5.3696 \pm 0.0024)$ GeV & $B_d$ & $1.37 \pm 0.14$ \\
$m_W$ & $(80.423 \pm 0.039)$ GeV & $\xi$ & $1.23 \pm 0.06$ \\
\hline
\end{tabular}
\caption{\small\sl Input parameters used in our fits. Values are taken from the
CKMfitter input table updated to summer 2006 \cite{CKMfitter}, with the exception of
$\xi$, for which we take the average performed in \cite{Hashimoto}. 
Unless otherwise stated, the first error is Gaussian, the second (where applicable) is uniform.} 
\label{tab:theo-input}
}

Turning to the MFV fits, the constraints used are collected in Table 
\ref{tab:UT-constraints} and the corresponding results are displayed in Figs. 
\ref{fig:MFV-fit1}-\ref{fig:MFV-fit2}. Some comments are in order on the choices of 
$|V_{ub}|$. Concerning the inclusive value, we mention that we tried {\em alternatively} 
all the averages reported in Ref. \cite{HFAG-winter06}. The latter are obtained by analyzing 
the inclusive data through use of three alternative theory prescriptions, namely BLNP
\cite{LNP,BLNP,BNP,Neubert1,Neubert2}, chosen in our final fits, and
\beqn
\begin{array}{ll}
|V_{ub}|^{\rm (incl)} = (44.6 \pm 2.0 \pm 2.0) \cdot 10^{-4}~, & \mbox{   HFAG average
using DGE \cite{DGE}}~,\\
[0.2cm]
|V_{ub}|^{\rm (incl)} = (50.2 \pm 2.6 \pm 3.7) \cdot 10^{-4}~, & \mbox{   HFAG average
using BLL \cite{BLL}}~.
\end{array}
\label{DGE-BLL}
\eeqn
The choice of the BLNP average is, for our purposes, the most conservative. In
fact, in the MFV fit II, it leads to two $R_t$ solutions which are closer to the SM one 
(eq. (\ref{SM-Rt-reference})) than in the cases where one uses either of the two other averages
in eq. (\ref{DGE-BLL}).

Concerning instead $|V_{ub}|^{\rm (excl)}$, other determinations are provided
e.g. by Refs. \cite{BallZwicky} and \cite{Abada}, also quoted in \cite{HFAG-winter06}. All
of them are consistent with each other and we took the result of \cite{MILC} for
definiteness.
\TABLE[!t]{
\begin{tabular}{clccccc}
\hline
Constraint & \hspace{8ex}Value &  Ref. & SM fit & MFV fit I & MFV fit II \\
\hline
$|V_{ud}|$ & $0.9738 \pm 0.0003$ & \cite{CKMfitter} & \checkmark & & \\
$|V_{us}|$ & $0.2257 \pm 0.0021$ & \cite{PDBook} & \checkmark & & \\
$\alpha$ & $(94 \pm 8)^\circ$ & \cite{UTfit} & \checkmark & & \\
$\gamma$ & $(83 \pm 19)^\circ$ & \cite{UTfit} & \checkmark & & \\
$\D M_{d}$ & $(0.507 \pm 0.004)$/ps & \cite{HFAG-winter06} & \checkmark & & \\
$\D M_{s}$ & $(17.77 \pm 0.12)$/ps & \cite{CDF-DMs} & \checkmark & & \\
$|\eps_K|$ & $(2.221 \pm 0.008)\times 10^{-3}$ & \cite{CKMfitter} & \checkmark & & \\
$\sin 2 \beta$ & $0.675 \pm 0.026$ & \cite{HFAG-winter06} & \checkmark & \checkmark & \checkmark \\
[0.1cm]
$|V_{cb}|^{\rm (incl)}$ & $(41.7 \pm 0.7) \cdot 10^{-3}$ & \cite{PDBook} & \checkmark & \checkmark & \checkmark \\
[0.1cm]
$|V_{cb}|^{\rm (excl)}$ & $(39.2 \pm 0.7 \pm 1.4) \cdot 10^{-3}$ & \cite{HFAG-winter06} & \checkmark & \checkmark & \\
[0.1cm]
$|V_{ub}|^{\rm (incl)}$ & $(45.2 \pm 1.9 \pm 2.7) \cdot 10^{-4}$ & \cite{HFAG-winter06} & \checkmark & \checkmark & \checkmark \\
[0.1cm]
$|V_{ub}|^{\rm (excl)}$ & $(35.5 \pm 2.5 \pm 5.0) \cdot 10^{-4}$ & \cite{MILC} & \checkmark & \checkmark & \\
\hline
\end{tabular}
\caption{\small\sl Constraints used in the various fits. The first error is Gaussian, the second
(where applicable) is uniform. A (\checkmark) indicates that the constraint is included in
the corresponding fit.} 
\label{tab:UT-constraints}
}

It can be noted that in the MFV fits of Table \ref{tab:UT-constraints} we did not include the 
constraints on $\alpha$ and $\gamma$. We mention that inclusion of $\gamma$ from the tree-level 
determination \cite{UTfit} in the MFV fit II has the effect of `selecting' the
large-$R_t$ solution (see below) among the two solutions displayed in Figs.
\ref{fig:MFV-fit1} and \ref{fig:MFV-fit2}, while inclusion of both $\alpha$ and $\gamma$ results in a single,
SM-like solution, with a quite large error. We decided to exclude the $\alpha$ and
$\gamma$ constraints from our main analysis, since our aim here is to make the tension
between a `large' $|V_{ub}|$ value and the present $\sin 2 \beta$ determination in
the context of MFV most transparent.

\FIGURE[!t]{
\includegraphics[width=0.60\textwidth]{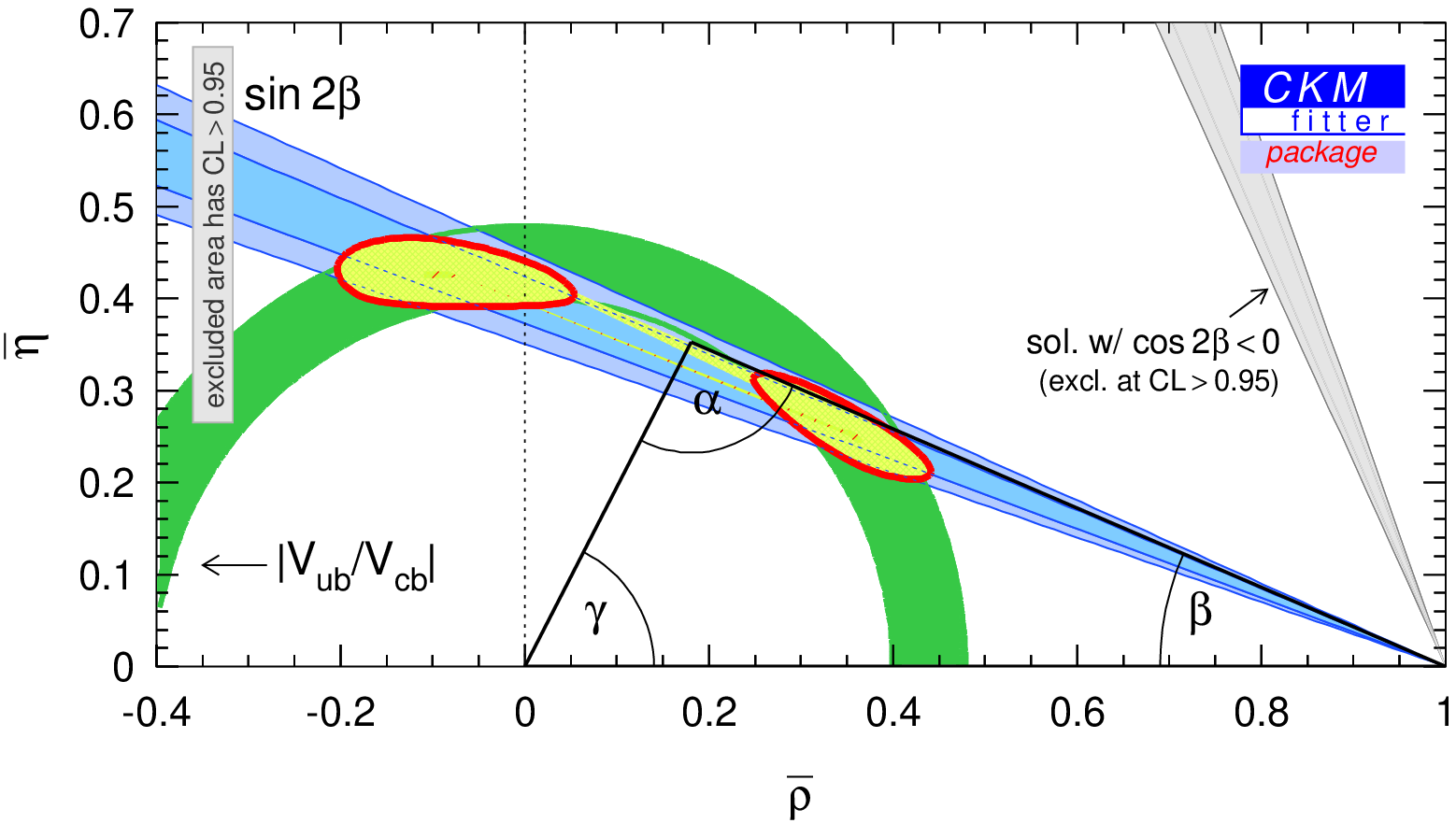}
\includegraphics[width=0.38\textwidth]{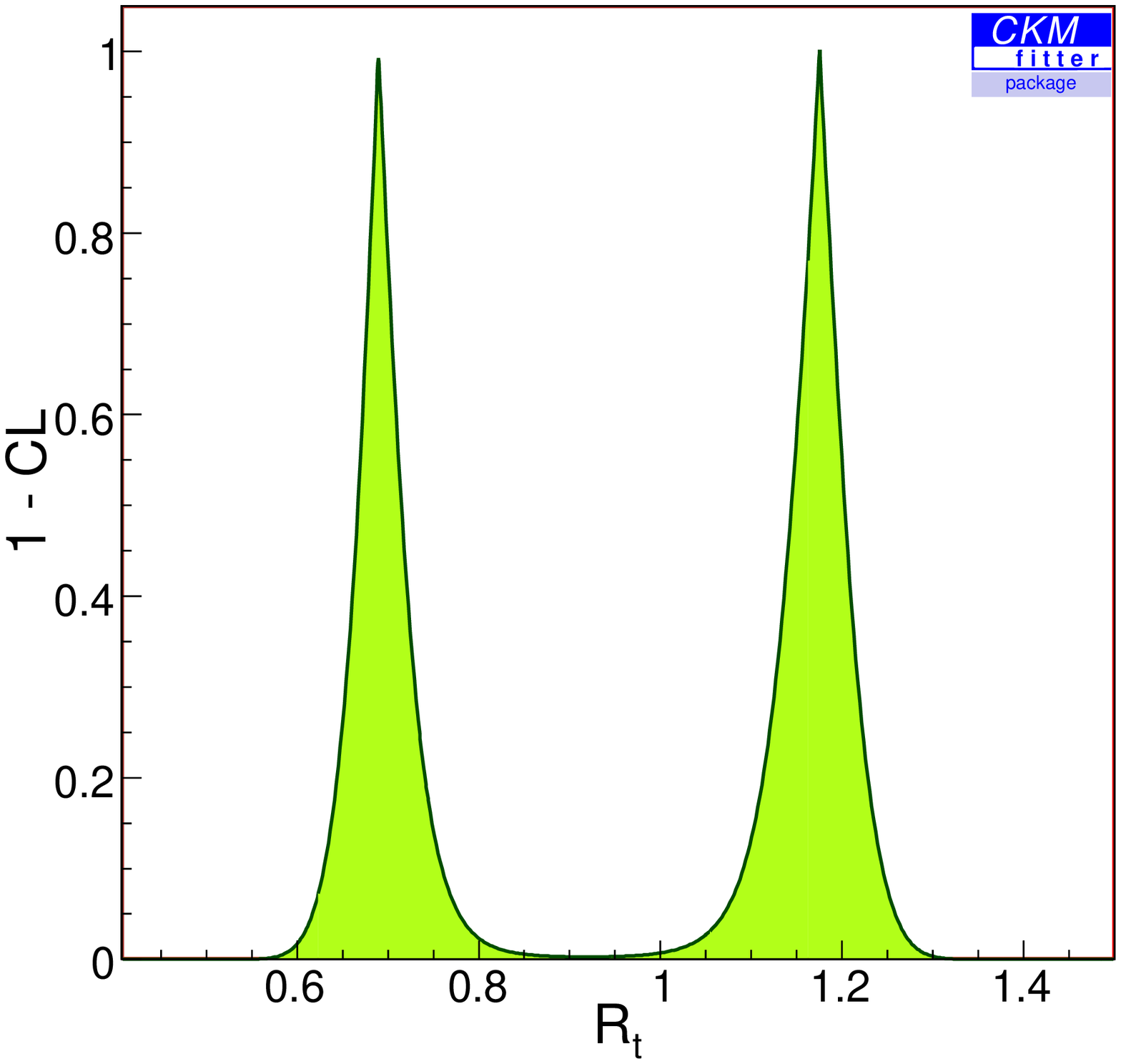}
\caption{\small\sl Results of the MFV fit I. See Table \ref{tab:UT-constraints} for the
constraints used. Left panel displays the selected area in the $(\ov \rho - \ov \eta)$
plane, and the UT from the SM fit (\ref{SM-Rt-reference}) is shown as reference. All contours are at 95\% CL. 
Right panel reports the corresponding confidence level profile for $R_t$.}
\label{fig:MFV-fit1}
}
\FIGURE[!t]{
\includegraphics[width=0.60\textwidth]{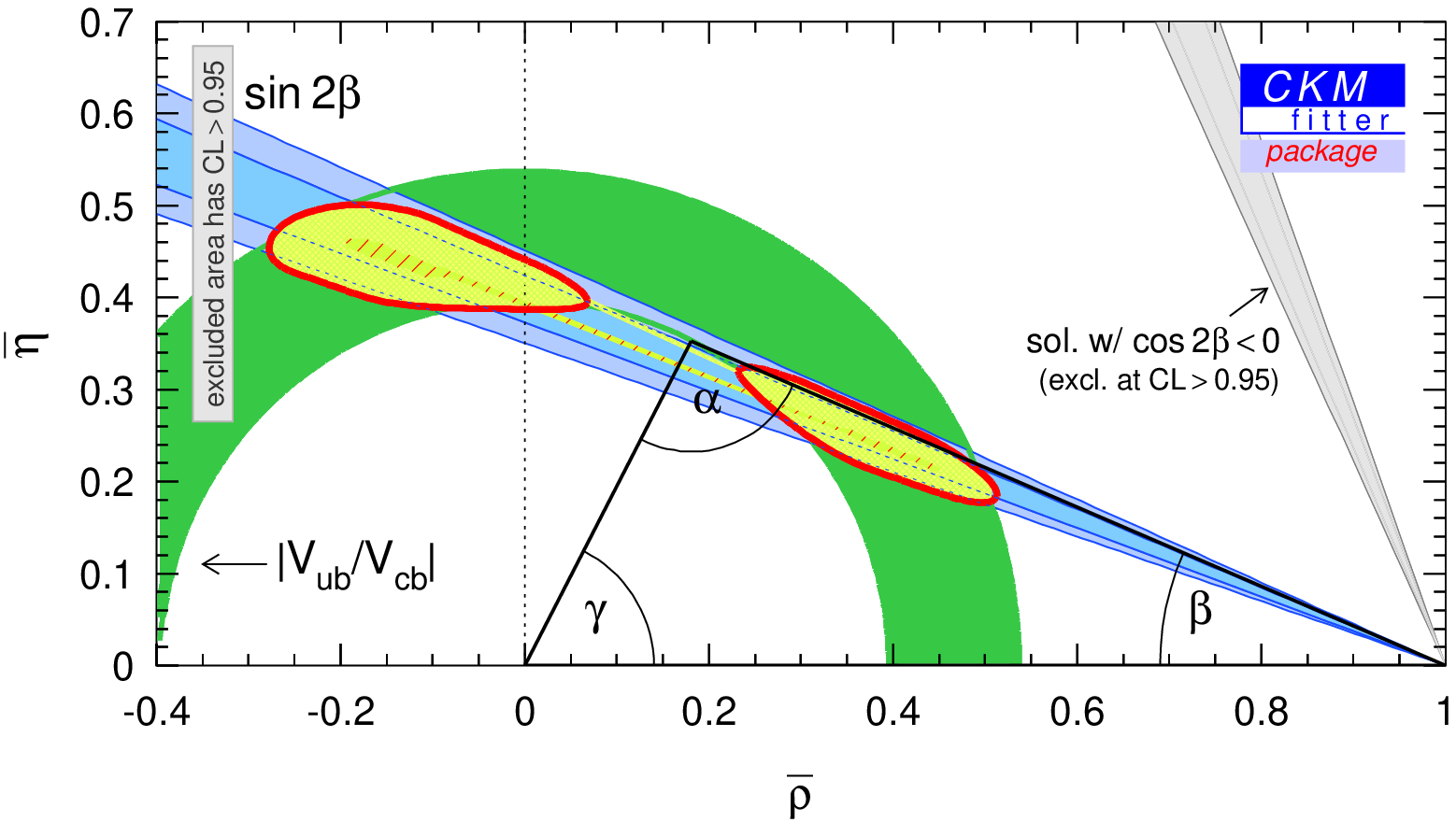}
\includegraphics[width=0.38\textwidth]{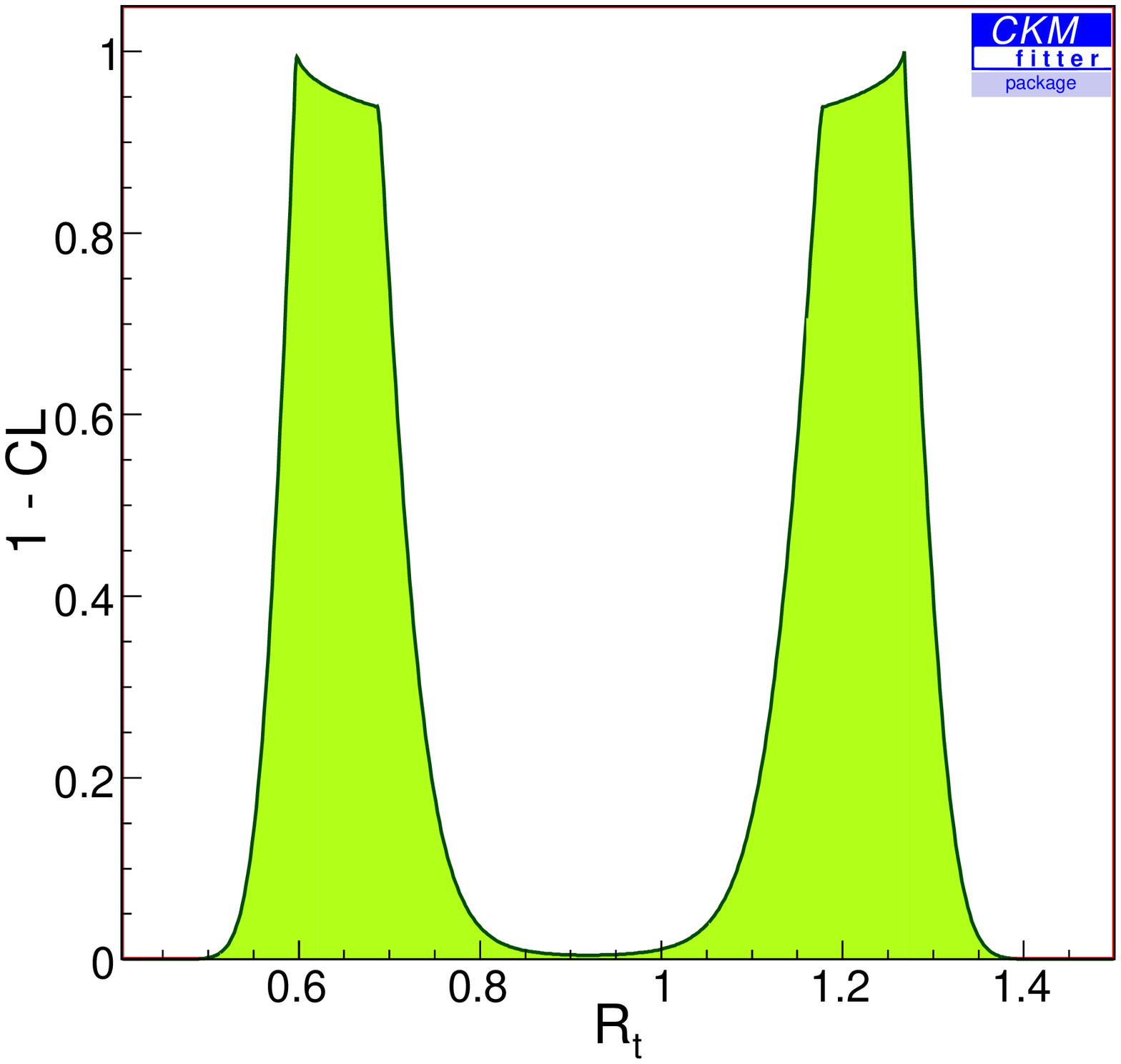}
\caption{\small\sl Results of the MFV fit II. See Table \ref{tab:UT-constraints} for the
constraints used. Left and right panels as in Fig. \ref{fig:MFV-fit1}.}
\label{fig:MFV-fit2}
}

The $R_t$ values resulting from the fits of Figs. \ref{fig:MFV-fit1}-\ref{fig:MFV-fit2} are
the following (95\% CL)
\beqn
\label{MFV1-Rt}
&(R_t^{(1)})_{\rm MFV-I} \in [0.62, 0.78]~, 
~~~(R_t^{(2)})_{\rm MFV-I} \in [1.07, 1.26]~,& \\
\label{MFV2-Rt}
&(R_t^{(1)})_{\rm MFV-II} \in [0.54, 0.79]~, 
~~~(R_t^{(2)})_{\rm MFV-II} \in [1.06, 1.34]~.& 
\eeqn
These values should be contrasted with the SM result for $R_t$, given in eq.
(\ref{SM-Rt-reference}). Some comments are in order here.
\begin{itemize}

\item In the MFV fit II, the inclusion of only $|V_{ub}|^{\rm (incl)}$ determines
two disjoint solutions for $R_t$ (eq. (\ref{MFV2-Rt})), which display a
discrepancy with respect to the SM solution (eq. (\ref{SM-Rt-reference})), at more than 
the 2$\sigma$ level.

\item The presence of two disjoint solutions for $R_t$ is featured also by the 
MFV fit I, where both the $|V_{ub}|^{\rm (incl)}$ and $|V_{ub}|^{\rm (excl)}$ determinations 
are included.

\item The two $R_t$ solutions turn out to be even better resolved in the MFV fit
I. This is due to the inclusion in the fit of the exclusive determinations for both $|V_{ub}|$ and
$|V_{cb}|$. In fact, on the one hand, the latter would shift the central value for the average 
of $|V_{ub}|/|V_{cb}|$ respectively down and up, as compared to the value obtained in the 
MFV fit II, and the two effects tend to cancel. 
On the other hand, the presence of two determinations for both $|V_{ub}|$ and $|V_{cb}|$ reduces 
the final $|V_{ub}|/|V_{cb}|$ error with respect to the MFV fit II case.

\end{itemize}

One possible approach to the above findings is to raise doubts on the $|V_{ub}|^{\rm (incl)}$
determination \cite{Ball-Vub-excl,Gambino} (see \cite{Ira} for the exclusive approach and 
\cite{Flynn-Nieves} for a very recent reanalysis).
We observe, however, that the inclusive value for $|V_{ub}|$ results from an average among quite 
a large number of modes, and the latter display consistency among each other. So the
average would definitely appear under control, were it not for the above mentioned (and well known)
discrepancy with the $|V_{ub}|$ value preferred by the global SM fit, close to the
exclusive determination.

In the following we will pursue the possibility that the $|V_{ub}|^{\rm (incl)}$
determination be the correct one, and answer the question whether the MFV MSSM can
account for either of the discrepancies $(R_t)_{\rm SM} - (R_t^{(1,2)})_{\rm MFV}$.
In particular, for $(R_t^{(1,2)})_{\rm MFV}$ we will take the results from the 
MFV fit II, which are the most conservative.
Our study addresses the eventuality that $|V_{ub}|$ should consolidate to a value 
{\em above} that required by the SM fit. Since such value should most 
probably lie somewhere in between the present exclusive and inclusive determinations, our study 
represents somehow the `limiting' case of a `high' $|V_{ub}|$ value.

We finally note that, since the MFV determination of the UT can rely only on a
handful of reasonably known observables (at present those in Table
\ref{tab:UT-constraints}), our study stresses the importance of an accurate determination
of $|V_{ub}|$ and of the angles $\alpha$ and $\gamma$. Angle determinations are fortunately a major task 
of the LHCb program and later, hopefully, of the Super-B \cite{SuperB}, which will also
allow a precision determination of $|V_{ub}|$.

\section[Corrections to $R_t$ within the MFV MSSM]{\boldmath Corrections to $R_t$ within the MFV
MSSM} \label{sec:MFV-MSSM}

\subsection[General formula for $R_t$]{\boldmath General formula for $R_t$}

In the previous Section, we have focused our attention on the determination of the
UT side $R_t$. This is because $R_t$ can in turn be expressed in terms of the $B_d$ and $B_s$ mass 
differences $\D M_d$ and $\D M_s$. For large $\tan \beta$, these quantities can undergo large SUSY corrections 
even in MFV \cite{BCRSbig}.
To display the `dependence'\footnote{The $R_t$ definition is of course only dependent on CKM matrix 
entries.} of $R_t$ on NP contributions to $\D M_{d,s}$, we write the following chain of equalities \cite{BBGT}, 
starting from the very $R_t$ definition
\beqn
R_t \equiv \frac{|V_{td} V_{tb}^*|}{|V_{cd} V_{cb}^*|} \cong \frac{1}{\la}\left|
\frac{V_{td}}{V_{cb}}\right| = \frac{\xi}{\la} \sqrt{\frac{m_{B_s}}{m_{B_d}}} 
\left( \sqrt{\frac{\D M_d}{\D M_s}} \right)_{\rm SM}~.
\label{Rt}
\eeqn
In presence of NP contributions to $\D M_{d,s}$, one can always write relations of the
kind
\beqn
\D M_q = \D M_q^{\rm SM} (1 + f_q)~, ~~~~q = d,s~,
\label{DMqNP}
\eeqn
where the l.h.s. represents the full {\em theoretical} predictions for the mass
differences, which must in turn be identified with the {\em experimentally} measured 
quantities, since we are supposing the presence of NP. Then, eq. (\ref{Rt}) can be rewritten 
as \cite{BBGT,BF,ABG}
\beqn
R_t &=& \frac{\xi}{\la} \sqrt{\frac{m_{B_s}}{m_{B_d}}} 
\sqrt{\frac{\D M_d}{\D M_s}} \sqrt{\frac{1+f_s}{1+f_d}} \nn \\
&=& 0.913 \left[ \frac{\xi}{1.23} \right] \sqrt{\frac{17.8/ {\rm ps}}{\D M_s}} 
\sqrt{\frac{\D M_d}{0.507/ {\rm ps}}} \sqrt{\frac{1+f_s}{1+f_d}}~,
\label{RtNP}
\eeqn
where in the last equality we have plugged in the experimental values for $\D M_{d,s}$.

In the light of eq. (\ref{RtNP}), we consider now the two solutions $R_t^{(1,2)}$ (see eq.
(\ref{MFV2-Rt})) from the MFV fit II of the previous Section. We want to address the question 
whether the discrepancy with respect to the SM solution can be accounted for
within the MFV MSSM. This will be the case if shifts $f_q$ from eq. (\ref{DMqNP}) are
sufficiently large to correct eq. (\ref{RtNP}) by the required amount.

\subsection[MFV MSSM with low $\tan \beta$]{\boldmath MFV MSSM with low $\tan \beta$}
\label{sec:MFV-low}

This case, corresponding to $\tan \be \simle 10$, was addressed in detail in Ref. \cite{ABG}. 
There it was found that corrections $f_q$ are positive and do not exceed a few percent.
It was also stressed that corrections tend to display alignment between the $\D M_d$
and $\D M_s$ cases, since NP contributions from scalar operators able to distinguish 
between the $q=d,s$ channels come with a factor of $m_b m_q / M_W^2$ and are
below the 1\% level. Therefore, within the MFV MSSM at low $\tan \beta$, the ratio 
$f_q = \D M_q^{\rm NP} / \D M_q^{\rm SM}$ tends to be the same for the $d,s$ cases and deviations 
on eq. (\ref{RtNP}) are not sufficient to reproduce the $R_t$ solutions of eq. (\ref{MFV2-Rt}).

\subsection[MFV MSSM with large $\tan \beta$]{\boldmath MFV MSSM with large $\tan \beta$}
\label{sec:MFV-high}

Within the MSSM at large $\tan \beta$, the formulae for $\D M_d$ and $\D M_s$ feature
also contributions from the so-called Higgs double penguins (DP), with exchange of $h_0, H_0$ 
and $A_0$ bosons. For values of $\tan \beta \simge 30$, and assuming MFV, the $H_0$ and $A_0$ 
double penguins usually provide the dominant NP contribution which is found to be {\em negative}
in the whole part of the SUSY parameter space with $M_A \in [200,500]$ 
GeV \cite{BCRS-NP,isidori-retico,BCRS-PL,BCRSbig}.
The proportionality of such contribution to the external quark masses makes it generically
negligible for the $\D M_d$ case, while for $\D M_s$ sizable corrections are possible.

In order to address the question whether the SUSY corrections to $\D M_d$ and $\D M_s$,
computed in the MFV MSSM at large $\tan \beta$, can produce the shifts required for $R_t$
to reach any of the solutions in eq. (\ref{MFV2-Rt}), we adopt the following strategy
\begin{itemize}
\item[\bf 1.] compute the SUSY contributions to $\D M_{d,s}$ in the MSSM with large $\tan \beta$;
\item[\bf 2.] perform the MFV limit, according to the EFT definition \cite{MFV};
\item[\bf 3.] study the quantity $(1+f_d)/(1+f_s)$ in the SUSY parameter space left after 
the MFV limit.
\end{itemize}

Concerning point {\bf 1}, the calculation can be accomplished following the procedure of
Ref. \cite{BCRSbig}, which allows for a resummation of large $\tan \beta$ corrections. For a
detailed description of this procedure, we refer to Section 2 of \cite{BCRSbig}. We mention
that in our numerical analysis we consistently take into account effects coming from
flavour off-diagonal squark mass matrices. We do not include, instead, effects arising from large 
$\tan \beta$ corrections to the Higgs propagator entering DP contributions. The latter
corrections were found to have a non-negligible effect only for $M_A \lesssim 160$ GeV
\cite{FGH}. Similar conclusions will be drawn in Ref. \cite{GJNT}. Since we do not
consider such light pseudoscalar Higgs masses in our numerical analysis, these new effects
do not change the basic findings of the present work.

Turning to point {\bf 2}, the MFV limit was performed in exact analogy with Ref.
\cite{ABG}. In particular we expand the soft terms, which enter the squark mass matrices, as 
functions of the SM Yukawa couplings. After such expansions, the soft terms are still parametrically
dependent on an overall normalization mass scale and on the dimensionless coefficients
tuning the proportionality to the Yukawa couplings themselves.

Finally, let us go to point {\bf 3}. From the above discussion, it is clear that the
SUSY parameters one needs to consider in the analysis are the same as in the low $\tan \beta$ 
case, with the addition of the neutral physical Higgs masses, entering the DP contributions. 
In order to have accurate numerical predictions for such masses, we used the package 
{\tt FeynHiggs} \cite{FeynHiggs1,FeynHiggs2,FeynHiggs3,FeynHiggs4}, which calculates the whole 
physical Higgs spectrum. The only additional parameter required by {\tt FeynHiggs} besides
those already present in the low $\tan \beta$ case is $M_{A}$. Consequently, in the notation 
of \cite{ABG}, parameters are
\beqn
\label{params-low}
&&\ov m~,~~ A~,~~ M_1~,~~ M_2~,~~ M_{\tilde g}~, \\
&&\mu~,~~ M_{A}~,~~\tan \beta~,
\label{params-high}
\eeqn
with the addition of the 12 `MFV coefficients', governing the proportionality of the soft
terms to the SM Yukawa couplings (see \cite{ABG}). We observe that all these parameters
are real, since the presence of complex phases would induce non-CKM CP violation in meson
mixings. The latter would then contradict the MFV hypothesis.

To explore the SUSY parameter space, we adopt the following strategy.
We study the ratio $(1+f_s)/(1+f_d)$, which enters $R_t$ in eq. (\ref{RtNP}), 
by generating the MFV coefficients with flat distributions in the same ranges as
\cite{ABG} (these ranges are also collected in Table \ref{tab:SUSY-pars}).

Concerning the parameters in eqs. (\ref{params-low})-(\ref{params-high}), we make the
following observation. For large $\tan \beta$, box contributions are in general much
smaller than the DP ones; in addition, the `alignment' of the NP box contributions between
the $\D M_{d,s}$ cases, stressed in Section \ref{sec:MFV-low}, is found to hold for large 
$\tan \beta$ as well. As a consequence, when DP contributions are small, one has 
$(1+f_s^{\rm box})/(1+f_d^{\rm box}) \approx 1$; conversely, when DP contributions are large, box 
corrections are just a small correction, with negligible modifications of the distribution shape.
\TABLE[!t]{
\begin{tabular}{p{5cm}ll}
\hline
Uniformly distributed & \hspace{1.7cm}Fixed \\
\hline
$a_{1,2,3} \in [0.25, 1]$ & $\tan \beta = \{30, 50 \}$ \\
[0.1cm]
$a_{4,5}, b_{1,...,8} \in [-1, 1]$ & $M_A = \{200, 500, 800 \}$ GeV \\
[0.1cm]
$A \in [-2, 2]$ TeV & $\overline m = 1$ TeV \\
[0.1cm]
$|\mu| \in [0.2, 2]$ TeV & $M_1 = M_2 = 500$ GeV \\
[0.1cm]
                         & $M_{\tilde g} = \{ 200, 500, 800 \}$ GeV \\
\hline
\end{tabular}
\caption{\small\sl Parameter choices made in our scans. For the definition of the 
MFV coefficients $a_i, b_j,$ as well as of the squark biliear and trilinear mass scales
$\overline m$ and $A$, see Ref. \cite{ABG}.} 
\label{tab:SUSY-pars}
}

For the reason above, and as already stressed in Section \ref{sec:MFV-low}, box
contributions alone are not able to account for the $R_t$ solutions given in eq.
(\ref{MFV2-Rt}). In order to address the analogous possibility for the double penguins, we set 
$\ov m = 1$ TeV, thus decoupling box contributions, which would partially
cancel the (dominant) DP corrections. Parameters $A$ and $|\mu|$ are instead generated
flatly in the ranges $A \in [-2,+2]$ TeV and $|\mu| \in [0.2,2]$ TeV. On the sign of $\mu$ we
will come back in the following discussion. The choice of the EW gaugino mass parameters 
plays a negligible role and we set them as $M_1 = M_2 = 500$ GeV. Finally, we considered
three discrete values for the gluino mass, namely $M_{\tilde g} = \{ 200, 500, 800 \}$
GeV. Our results turn out to bear negligible dependence also on the choice of the gluino
mass: we will further comment on this in due course. In our plots, the value $M_{\tilde g}
= 500$ GeV is chosen for definiteness.
\FIGURE[!t]{
\includegraphics[width=0.49\textwidth]{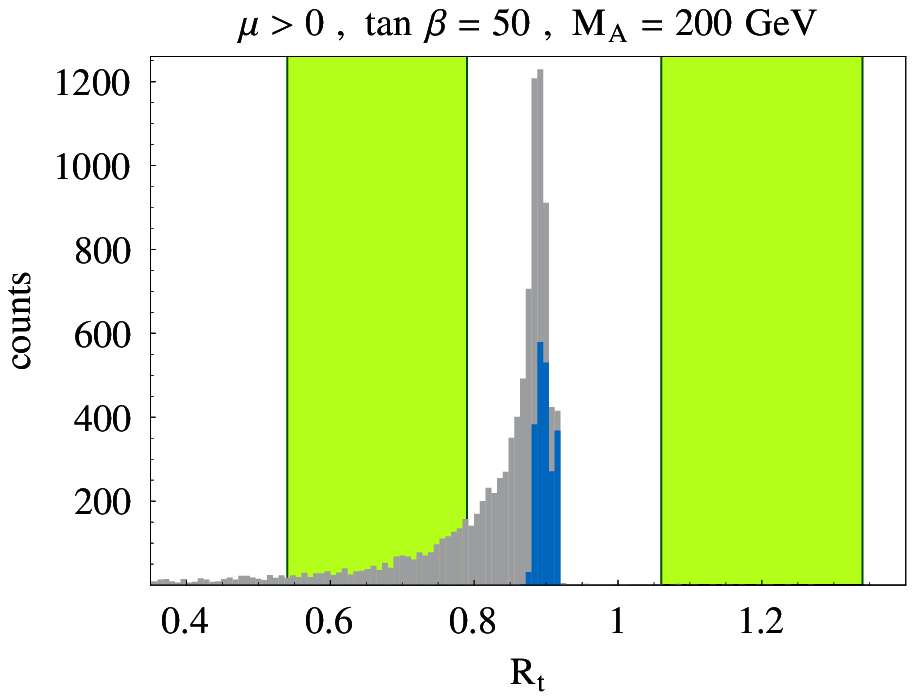}
\includegraphics[width=0.49\textwidth]{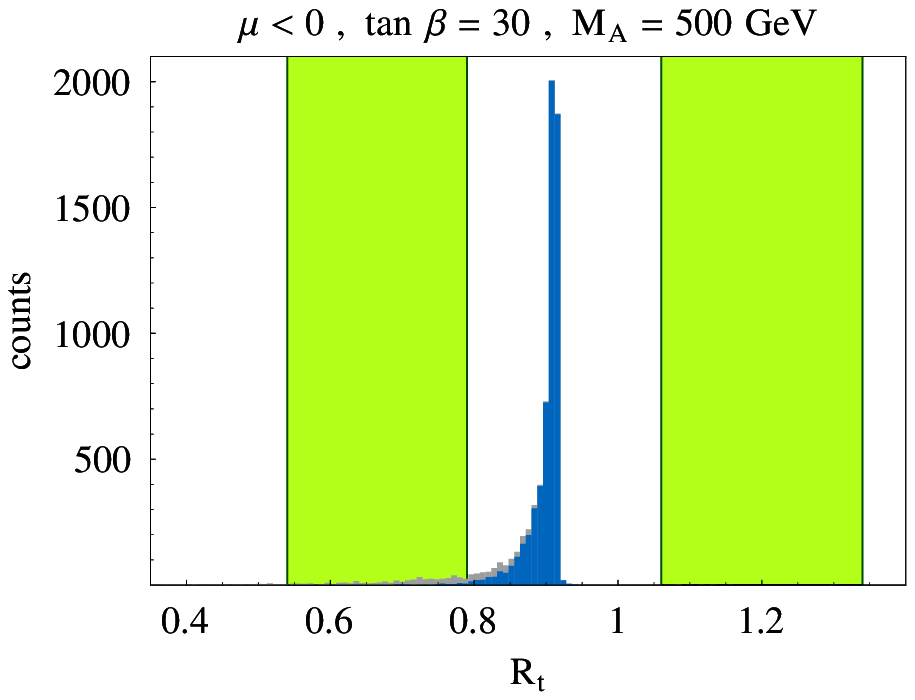}
\caption{\small\sl Distributions of values for $R_t$, resulting after scanning $A$, $|\mu|$
and the MFV parameters (see text for details). The left panel corresponds to $\mu > 0$,
$M_{A} = 200$ GeV and $\tan \beta = 50$; the right panel corresponds to $\mu < 0$,
$M_{A} = 500$ GeV and $\tan \beta = 30$. The grey (blue) distributions are without
(with) the constraint (\ref{Bsmumu-exp}). The vertical green bands correspond to the $R_t$ solutions given 
in eq. (\ref{MFV2-Rt}).}
\label{fig:Rt}
}

On the other hand, the quantity $(1+f_s)/(1+f_d)$ bears obviously a strong dependence on the choice
of the mass $M_{A}$ (which also sets the scale for the $H_0$ DP) and on the value of
$\tan \beta$. We have explored, in discrete steps, the ranges $M_A \in [200,800]$ GeV and 
$\tan \beta \in \{30,50\}$.

This completes the description of the choices made for the various SUSY parameters in our MonteCarlo 
study.\footnote{We mention that, strictly 
speaking, our choice of SUSY masses does not always satisfy the condition $M_{\rm SUSY} \gg M_A$, 
required by the effective Lagrangian
framework of Refs. \cite{BCRSbig,isidori-retico}. We assume, as also done in the existing
literature, that the latter approach be applicable also for heavy Higgs masses comparable
in size with $M_{\rm SUSY}$. A fully rigorous treatment would require a two loop
calculation in SUSY, beyond the scope of the present work.} 
These choices are also summarized in Table \ref{tab:SUSY-pars}. In Fig. \ref{fig:Rt} we show as grey 
distributions the implied effects on $R_t$ for choices of $M_A$ and $\tan \beta$ where such effects are significant. 
The left panel corresponds to the choice $M_A = 200$ GeV, $\tan \beta = 50$ and $\mu > 0$, while the right panel 
corresponds to $M_A = 500$ GeV, $\tan \beta = 30$ and $\mu < 0$.

Here we note that, from the left panel of Fig. \ref{fig:Rt}, the long tail looks able to reproduce the lower $R_t$ 
solution given in eq. (\ref{MFV2-Rt}). However, a large portion of the tail is actually unphysical, since it 
corresponds to values of $\D M_s$ which are outside the range allowed by the theoretical error ($\D M_d$ is 
only weakly affected by NP contributions, as we saw at the beginning of this section). When calculating $R_t$ 
out of $\D M_d^{\rm MSSM}$ and $\D M_s^{\rm MSSM}$, one should in fact take into account
the constraints coming from the $\D M_q$ measurements. However,
these constraints are by far dominated by the corresponding theoretical error.
We stress that, in our case, the latter must take into account not only the uncertainty coming 
from the lattice parameters, but also the error associated with the CKM entries. For these
entries one must in fact consistently use the determinations coming from the MFV fit, whose 
associated error is larger than that resulting from the usual SM fit determination.
\FIGURE[!t]{
\includegraphics[width=0.49\textwidth]{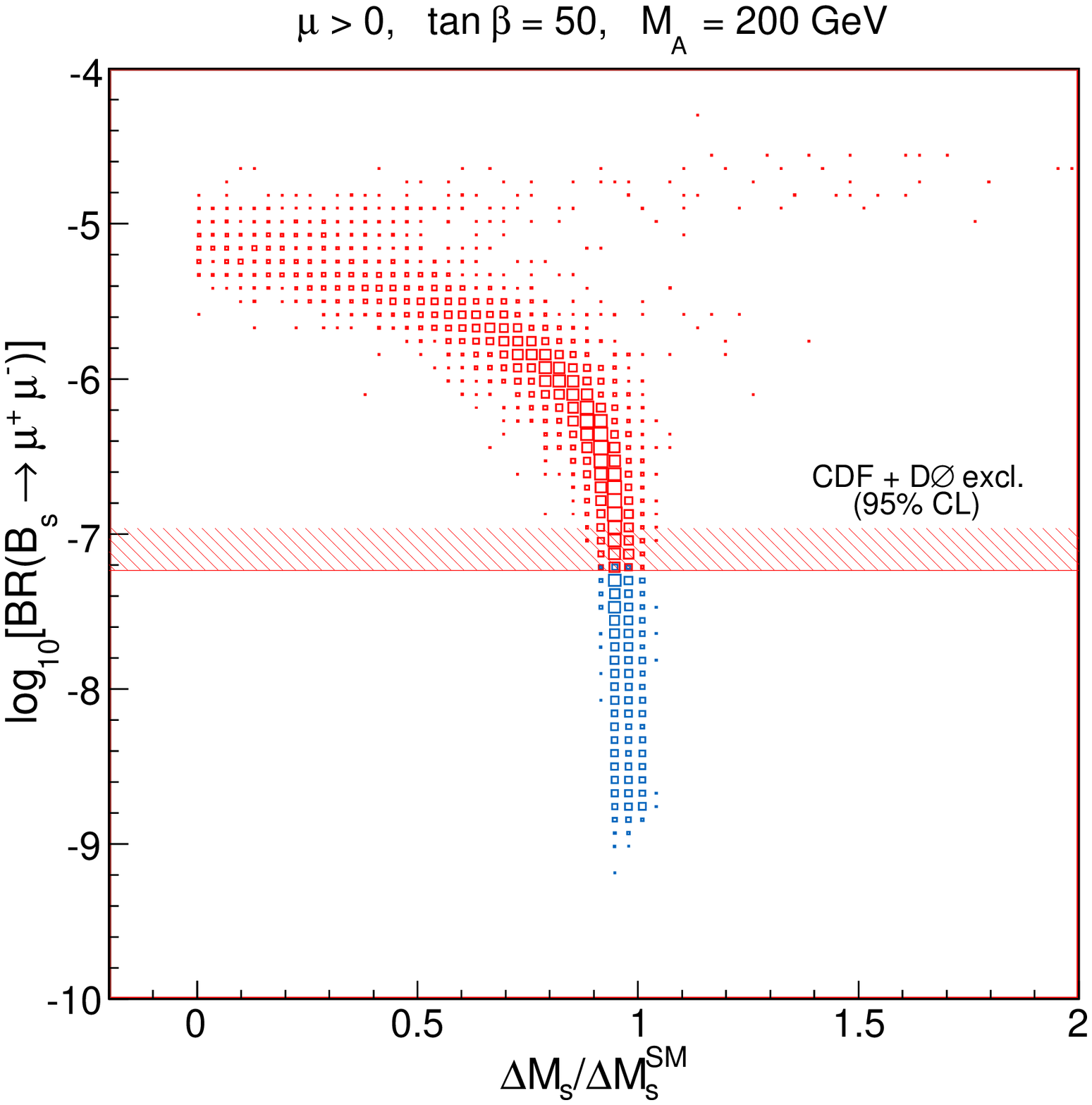}
\includegraphics[width=0.49\textwidth]{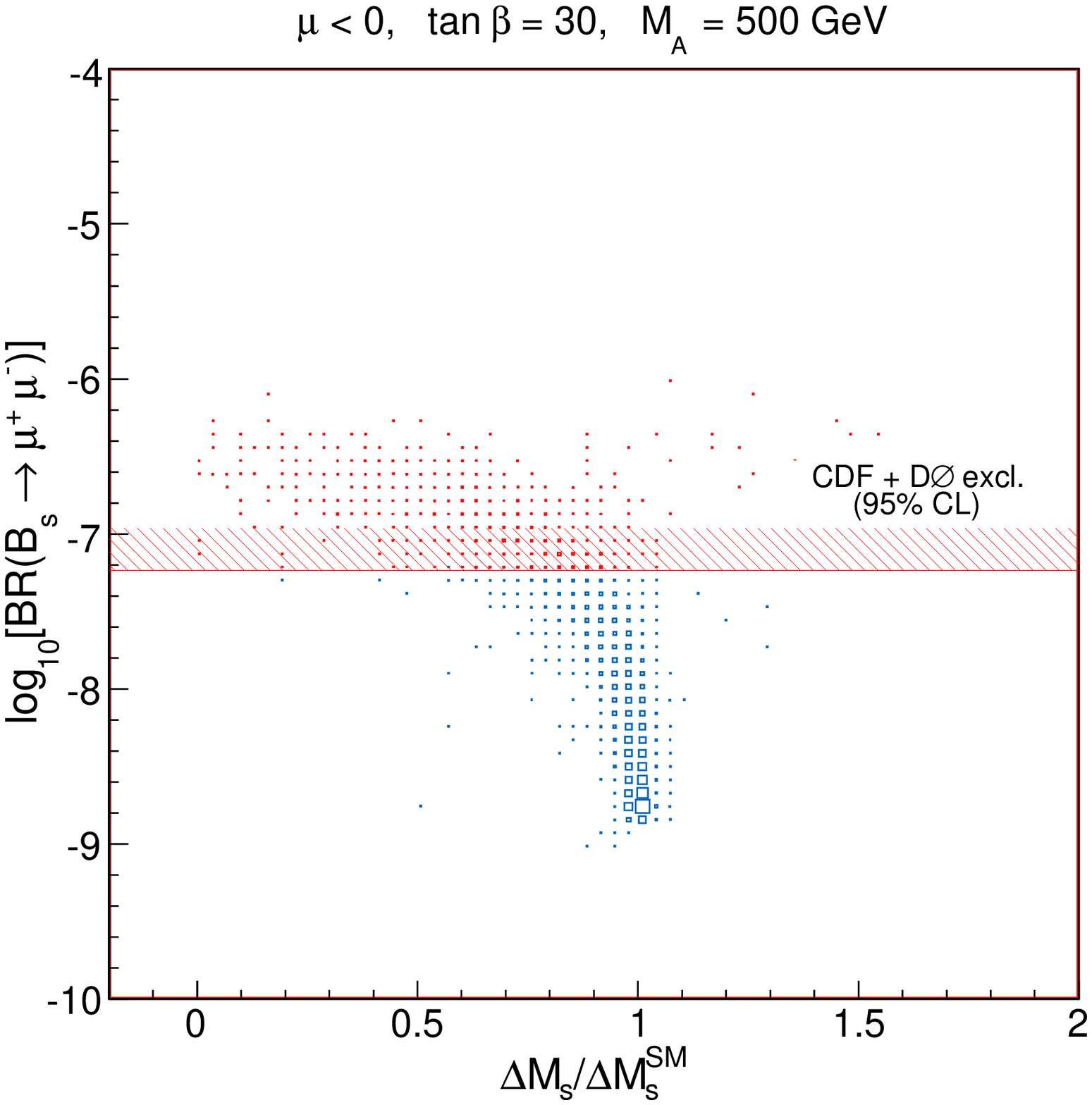}
\caption{\small\sl Correlation between BR$(B_s \to \mu^+ \mu^-)$ and $\D M_{d,s}$ in 
the MFV MSSM at large $\tan \beta$. Parameters as in Fig. \ref{fig:Rt}.}
\label{fig:Bsmumu-DMs}
}

The largeness of the final theoretical error makes then the $\D M_q$ constraints
not very effective. We decided not to include this `filter', since a much more effective one is
provided by the BR$(B_s \to \mu^+ \mu^-)$ upper bound\footnote{The BR$(B_s \to \mu^+ \mu^-)$ was 
calculated following exactly the same procedure as the one adopted for meson mixings
\cite{BCRSbig}. For a recent discussion of these observables using \cite{BCRSbig}, although in a 
different context, see Ref. \cite{CMW}.}. The latest combined bound from the CDF and D{\O}
collaborations reads \cite{CDF+D0-bsmumu}\footnote{For the previous bound from the CDF
collaboration, see \cite{CDF-bsmumu}.}
\beqn
{\rm BR}(B_s \to \mu^+ \mu^-) < 5.8 \times 10^{-8}~,~~~~\mbox{[95\% CL]}~.
\label{Bsmumu-exp}
\eeqn
The distribution of values for $R_t$ which survives the $B_s \to \mu^+ \mu^-$ constraint is 
shown in blue in Fig. \ref{fig:Rt}. For positive $\mu$, the constraint completely excludes the 
possibility of significant corrections to $R_t$, and sets the latter back to a SM-like
value. For negative $\mu$ a similar conclusion holds, considering that less than 1\% of 
the MC-generated values for $R_t$ lie within the 2$\sigma$ band of the lower $R_t$
solution, after the constraint (\ref{Bsmumu-exp}) has been taken into account.
We note that our results in Fig. \ref{fig:Rt} can be directly compared
with Fig. 3 of \cite{FGH}, which reports the quantity 
$R_{sd} = (\D M_s/\D M_s^{\rm SM})/(\D M_d/\D M_d^{\rm SM})$. The latter is related to
$R_t$ through $R_t = 0.913 \times \sqrt{R_{sd}}$. One can see that our MonteCarlo allows
slightly larger maximal effects than in the case of \cite{FGH}. This is most likely due to
the fact that we do not include the constraints from $\ov B \to X_s \gamma$ and $B \to
\tau \nu$, in constrast to Ref. \cite{FGH}. Including these constraints would however only
strenghten our conclusions.

The severeness of the constraint (\ref{Bsmumu-exp}) on the allowed MFV MSSM corrections
to $\D M_s$ is shown in Fig. \ref{fig:Bsmumu-DMs}. The latter reports the correlation 
\cite{BCRS-PL} between BR$(B_s \to \mu^+ \mu^-)$ and $\D M_s$ in the MFV MSSM\footnote{For
a discussion of the correlation in the general MSSM we refer the reader to \cite{Foster}.} 
at large $\tan \beta$, that is roughly given by
\beqn
\D M_s^{\rm DP} \propto 
- \left( \frac{M_A}{\tan \beta} \right)^2 {\rm BR}(B_s \to \mu^+ \mu^-)~.
\label{DMs-BR}
\eeqn 
An interesting aspect of the above correlation is that, in the regime of DP
dominance, a large part of the dependence on SUSY parameters other than $\tan \beta$ and
$M_A$ {\em cancels}, since it is common to BR$(B_s \to \mu^+ \mu^-)$ and $\D M_s$. This is
in particular true of the `$\eps$-factors' $\eps_Y, \tilde \eps_3$ and $\eps_0$ (compare
eqs. (6.25) and (6.40) of Ref. \cite{BCRSbig}), which incorporate the 
dependence on the SUSY particles entering the loops. This explains, for example, the
negligible dependence of the correlation (\ref{DMs-BR}) on the choice of the gluino mass,
already mentioned in the discussion of the parameter ranges for the MonteCarlo. A
similar finding holds for the $R_t$-distributions of Fig. \ref{fig:Rt}, once the 
$B_s \to \mu^+ \mu^-$ bound is taken into account.

In Fig. \ref{fig:Bsmumu-DMs}, we show two representative cases for the correlation
(\ref{DMs-BR}). As stated previously in the text, parameters are chosen in order to have large
(correlated) effects. For positive $\mu$, Fig. \ref{fig:Bsmumu-DMs} (left panel) shows
that the bound (\ref{Bsmumu-exp}) basically excludes effects on $\D M_s$ exceeding $- 10$\%.
In addition, the correlation with the prediction on $B_s \to \mu^+ \mu^-$ is largely
lost in the allowed region.

For negative $\mu$, neutral Higgs contributions to both BR$(B_s \to \mu^+ \mu^-)$ and $\D M_s$ 
are generically larger than in the corresponding positive $\mu$ case. Furthermore,
according to eq. (\ref{DMs-BR}), one can choose larger values for $M_A$ and/or lower
values for $\tan \beta$ in order to have a large correction to $\D M_s$ while fulfilling the 
constraint on BR$(B_s \to \mu^+ \mu^-)$.\footnote{We thank P. Paradisi for drawing this 
point to our attention.} In Fig. \ref{fig:Bsmumu-DMs} (right panel) we report a case with these
features, corresponding to $\tan \beta = 30$ and $M_A = 500$ GeV.  In fact, for negative 
values of $\mu$, one needs $M_A \gtrsim 500$ GeV if $\tan \beta \gtrsim 30$ in order to observe a 
significant correlated effect between $\D M_s$ and BR$(B_s \to \mu^+ \mu^-)$. One should also keep 
in mind that for $\mu < 0$ the MSSM worsens the $(g-2)_{\mu}$ discrepancy with respect to the 
SM \cite{isidori-paradisi,IMPT}.

\FIGURE[!t]{
\includegraphics[width=0.49\textwidth]{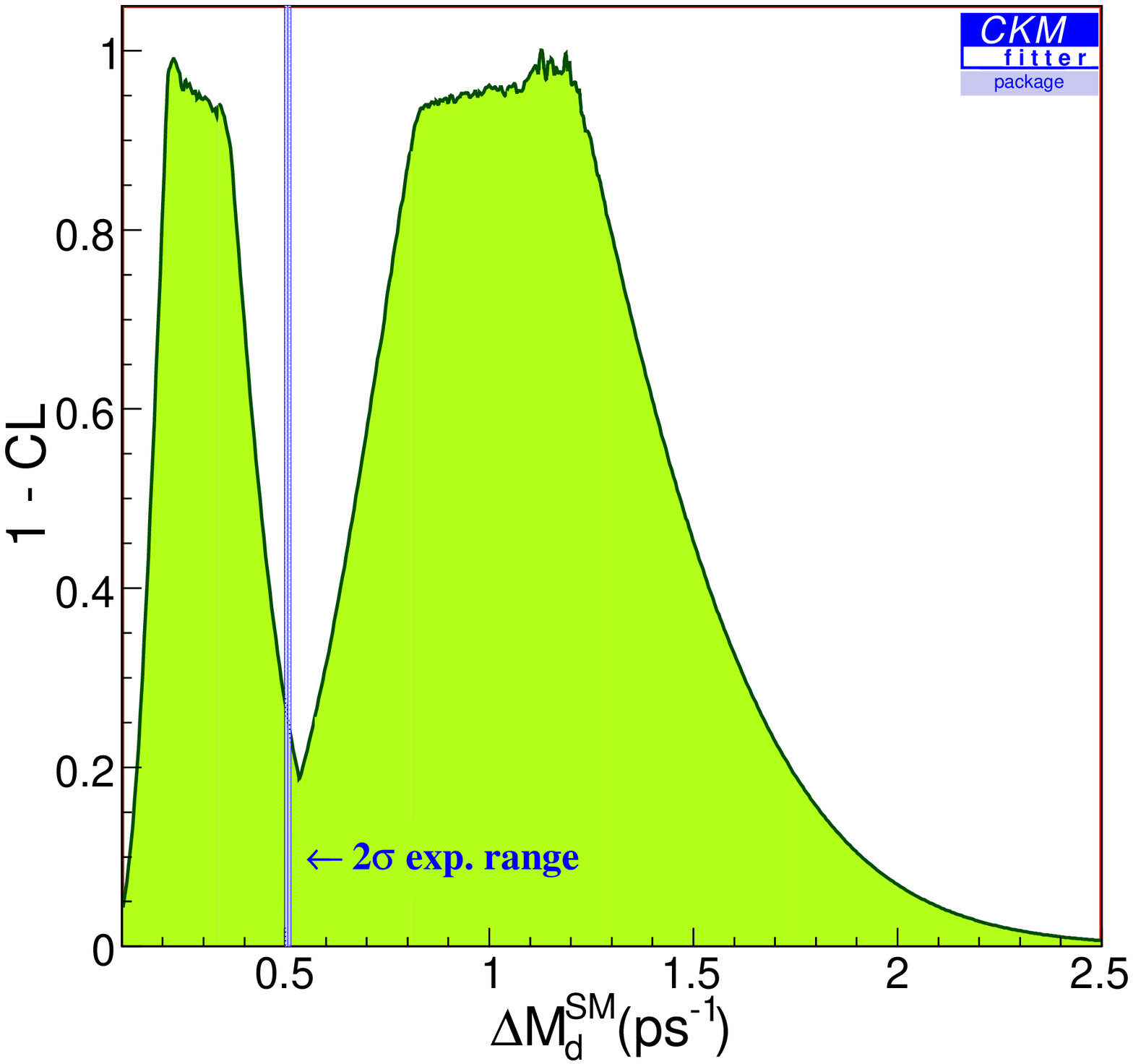}
\includegraphics[width=0.49\textwidth]{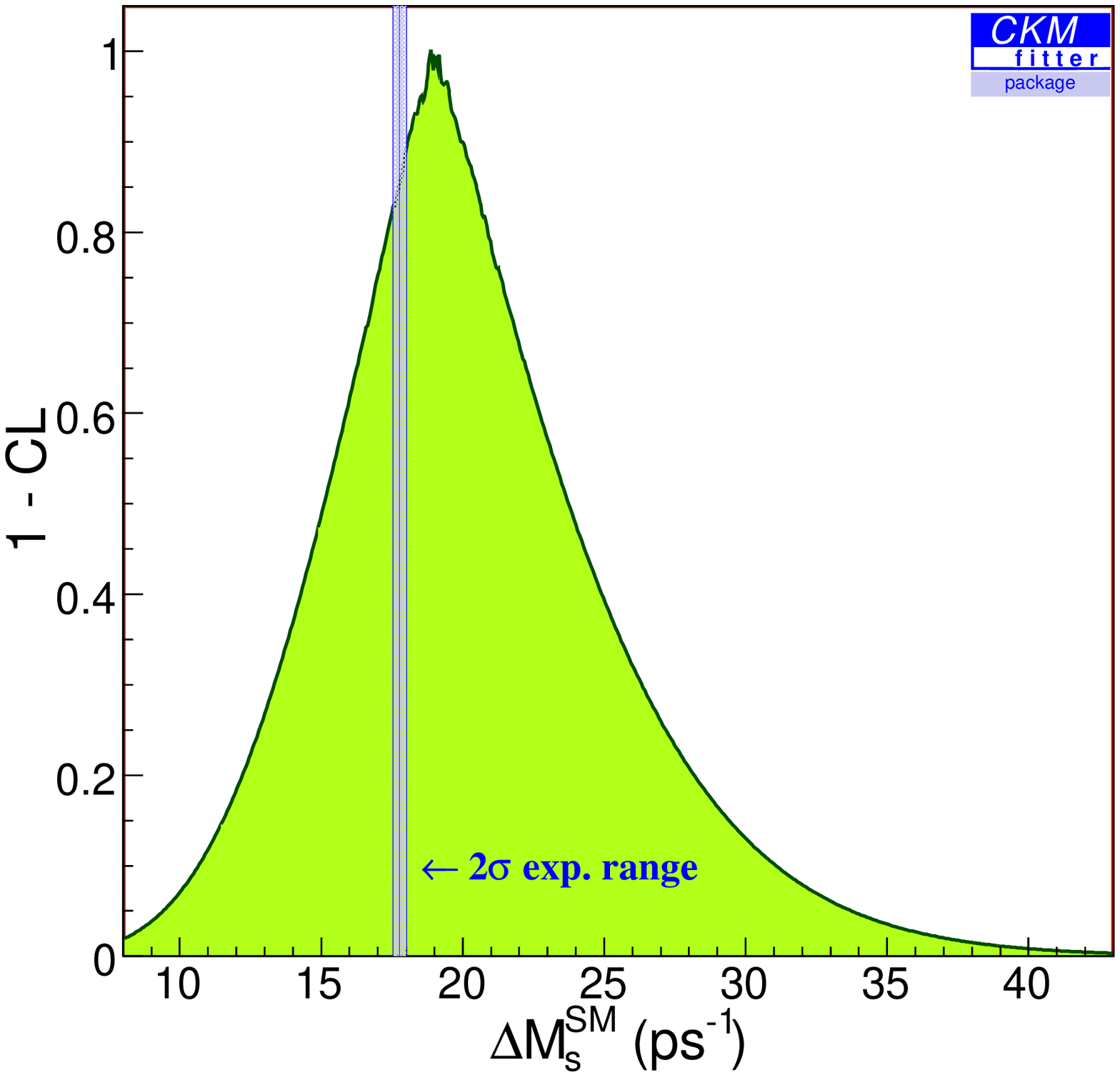}
\caption{\small\sl CL profiles for $\D M_d^{\rm SM}$ (left panel) and $\D M_s^{\rm SM}$
(right panel) for the MFV fit II.}
\label{fig:DMq-SM}
}
We note here some peculiar features of the MFV MSSM corrections to $\D M_{d,s}$, which
make them unable to reproduce any of the two $R_t$ solutions (\ref{MFV2-Rt}).
\begin{itemize}
\item[{\bf a)}] Among the CKM elements entering the SM formulae for $\D M_{d,s}$, only 
$V_{td}$ is significantly modified within the MFV fit with respect to $V_{td}^{\rm SM}$. Since $V_{td}$ 
enters only $\D M_d$, in order to reproduce the experimental values one would need a correspondingly
large correction $f_d$ on $\D M_d^{\rm SM}$ and a negligible $f_s$ correction to 
$\D M_s^{\rm SM}$. The MFV MSSM tends instead to give large corrections only
to $\D M_s$. In fact DP contributions are sensitive to the external quark masses, and NP effects 
scale as $m_q m_b$ for $\D M_q$.\footnote{In the light 
of this, a model able to produce large corrections to $\D M_d$ while leaving those on 
$\D M_s$ negligible looks quite hard to construct. An interesting case is however that
studied by \cite{FGH,GJNT}.}
\item[{\bf b)}] In absence of the BR$(B_s \to \mu^+ \mu^-)$ constraint, corrections to $R_t$
tend to reproduce the lower solution, since $f_s$ in eq. (\ref{RtNP}) is negative.
On the other hand, the higher solution looks the favored one, \eg by the tree-level
determination for the angle $\gamma$.
\end{itemize}
Concerning point {\bf a}, we show in Fig. \ref{fig:DMq-SM} the CL profiles for 
$\D M_{d,s}^{\rm SM}$ within the MFV fit II. As one can see, in this case the SM formula for
$\D M_d$ favors two solutions, corresponding to the shifts in the values of $V_{td}$. On
the other hand, $\D M_s^{\rm SM}$ is perfectly compatible with the experimental result.

We have studied the distribution of values for $R_t$ as given in eq.
(\ref{RtNP}) for different choices of the relevant low-energy parameter $\xi$, which brings 
the largest contribution to the overall error.\footnote{We mention here that the
distributions of values for $R_t$ shown in Fig. \ref{fig:Rt} are calculated for
$\xi$ set to its central value. Taking into account the $\xi$ error does not change the
relevant features of the distributions.} Results are displayed in the left panels of Fig.
\ref{fig:Rt-gamma-vs-xi}. In addition, when assuming MFV, a distribution of values for $R_t$ 
can be translated into a corresponding distribution for $\gamma$, with the only additional
uncertainty of $\sin 2 \beta$, to which however $\gamma$ is only weakly sensitive.
Then, similarly to $R_t$, one can study the dependence of the $\gamma$ determination on
the value of $\xi$. Results are shown in the right panels of Fig. \ref{fig:Rt-gamma-vs-xi}. 
The two solutions for $\gamma$ corresponding to the $R_t$ determinations of eq.
(\ref{MFV2-Rt}) are as follows (95\% CL)
\beqn
(\gamma^{(1)})_{\rm MFV-II} = [21,48]^\circ~,~~~(\gamma^{(2)})_{\rm MFV-II} =
[87,120]^\circ~.
\label{MFV2-gamma}
\eeqn
They are displayed in Fig. \ref{fig:Rt-gamma-vs-xi} (right) as horizontal green bands.
\FIGURE[!t]{
\includegraphics[width=0.49\textwidth]{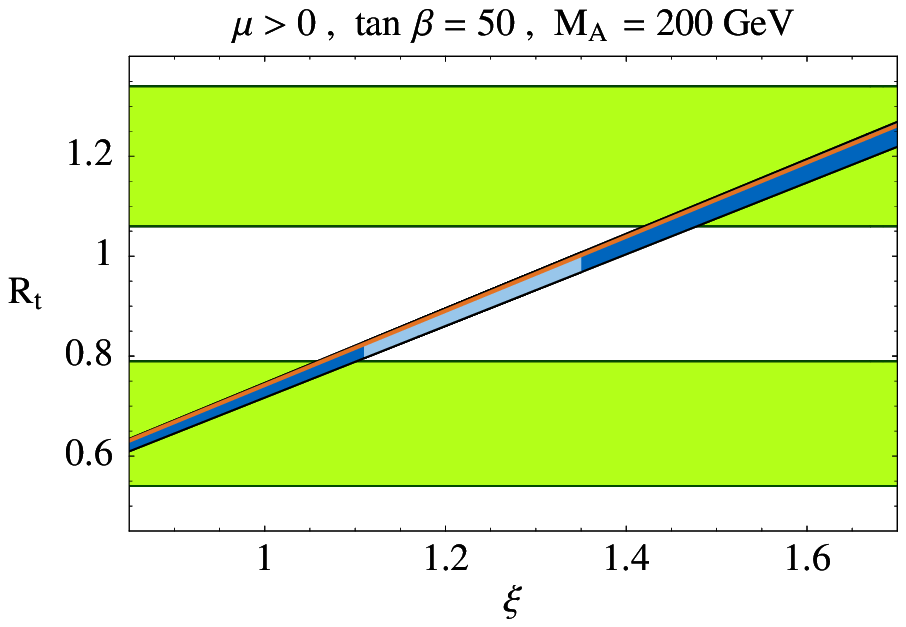}
\includegraphics[width=0.49\textwidth]{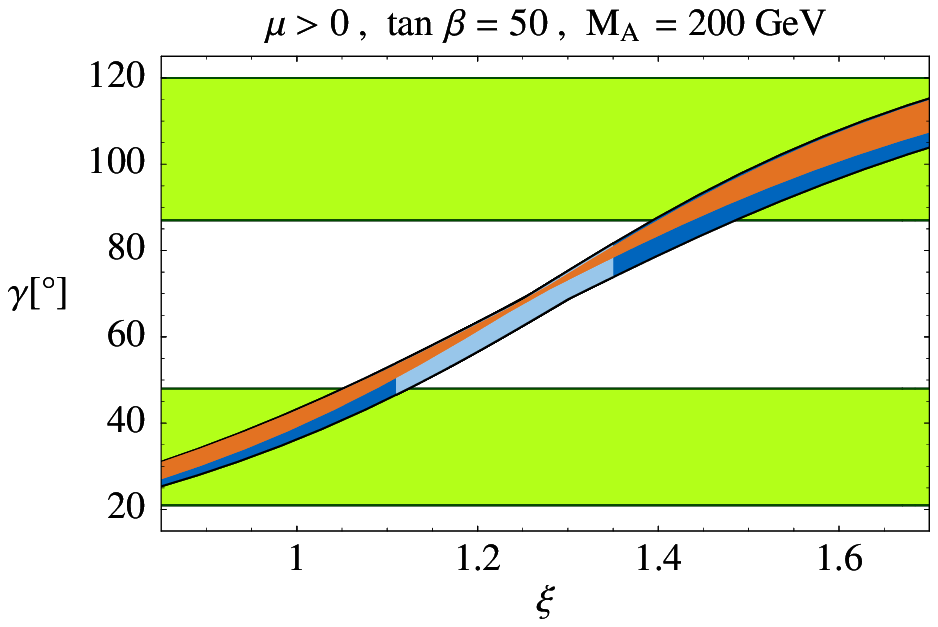}\vspace{0.4cm}
\includegraphics[width=0.49\textwidth]{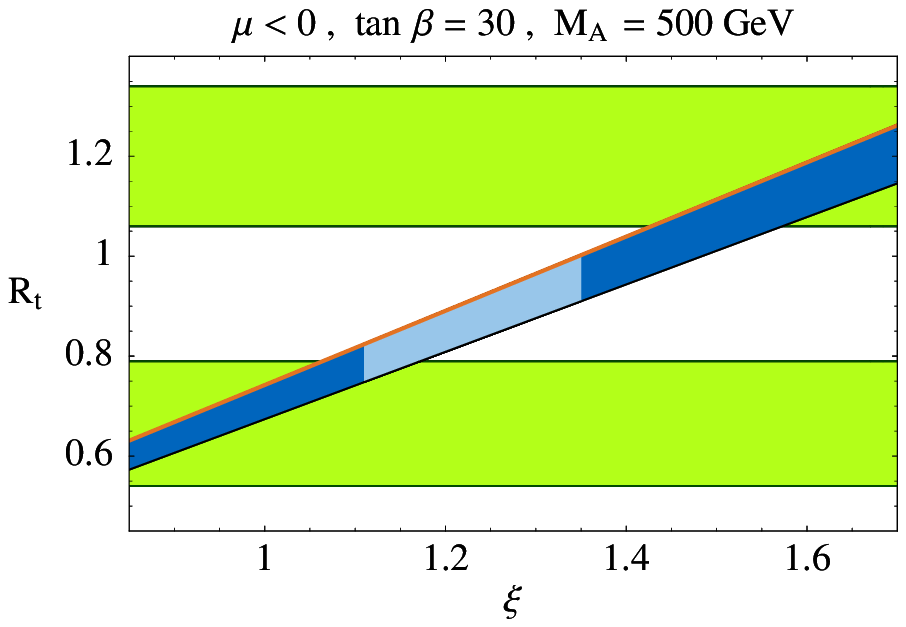}
\includegraphics[width=0.49\textwidth]{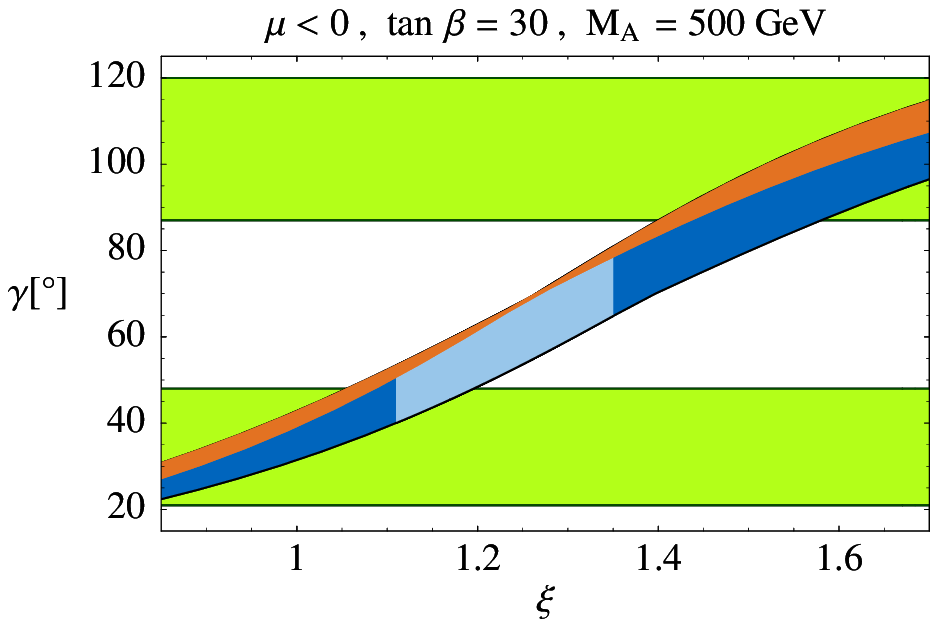}
\caption{\small\sl $R_t$ vs. $\xi$ and $\gamma$ vs. $\xi$ plots for $\mu > 0$ (high
panels) and $\mu < 0$ (low panels). The blue bands represent the 
range of values allowed within the MFV MSSM. 
The central superimposed light blue belts correspond to the present 2$\sigma$-range 
for $\xi$ from lattice QCD (see Table \ref{tab:theo-input}). The orange lines report the SM 
results. Finally the two horizontal green bands represent the $R_t$ (left panels) or $\gamma$ 
(right panels) solutions from the MFV fit II, eqs. (\ref{MFV2-Rt}) and (\ref{MFV2-gamma}).}
\label{fig:Rt-gamma-vs-xi}
}

We conclude this Section by noting that further support to the above results is provided
by studying the quantity $\eps_K$. The latter presents features exactly analogous to $\D
M_d$. In particular: (i) given its dependence on $V_{td}$, in the MFV fits the SM
formula gives two solutions; (ii) to reproduce the experimental value, one would need
relatively large NP contributions; (iii) within the MFV MSSM, such contributions are
however negligible in the full parameter space, so that agreement would again require the
parameter $B_K$ to be substantially different ($> 2 \sigma$) from the present lattice
determination.

\section{Conclusions}

In the present paper we have pointed out that the large value of $|V_{ub}|$ from
inclusive tree-level determinations is not only a challenge for the Standard
Model, but also for the MSSM with MFV. The patterns and the size of modifications of 
$\Delta M_d$ and $\Delta M_s$ in the MSSM with MFV do not allow a consistent description
of the data in this case, unless very significant modifications of 
non-perturbative parameters determinations relative to the lattice ones are made.
Concerning in particular the modifications allowed to $\D M_s$, we conclude that scenarios with 
200 GeV $\lesssim M_A \lesssim$ 500 GeV and $\tan \beta \simeq 50$ that predict a significant 
suppression for $\D M_s$ in {\em correlation} with an enhancement for BR$(B_s \to \mu^+ \mu^-)$ 
have to be fine-tuned in order not to violate the new combined bound on the latter decay mode 
from the CDF and D{\O} collaborations. Relatively large correlated effects can however still 
occur for negative values of $\mu$ and large values for $M_A \gtrsim 500$ GeV, increasing with 
increasing $\tan \beta \gtrsim 30$.

The resolution of the $|V_{ub}|$ problem -- that is, the discrepancy between the exclusive
and inclusive determinations of this quantity -- calls in the first place for a better
theoretical control on the relevant hadronic quantities involved in either determination.
If the two determinations should then agree on a central value in the ballpark of the
present inclusive average, insight on the tension existing within the SM and the MFV MSSM 
will require a facility like Super-B \cite{SuperB}, where also a very precise determination of 
the angle $\gamma$ from tree-level processes is possible. As we have seen, the precise knowledge 
of this angle could definitely tell us whether MSSM with MFV has a chance to be a correct 
description of flavour violating processes.

\section*{Acknowledgments}

We thank the authors of Ref. \cite{GJNT} for communicating to us the general pattern of
the large $\tan \be$ corrections studied in their work. We warmly acknowledge S. T'Jampens 
for kind feedback on the interpretation of the {\tt CKMfitter} package output and very useful remarks. 
Thanks are also due to J. Charles, A. Hoecker and V. Tisserand for useful correspondence
as well as U. Haisch, F. Mescia and P. Paradisi for important remarks. Finally, D.G.
acknowledges G. D'Agostini and M. Pierini for  useful discussions. 
This work has been supported in part by the Cluster of Excellence ``Origin and Structure of the 
Universe'' and by the German `Bundesministerium f{\"u}r Bildung und Forschung' under contract 
05HT6WOA. D.G. also warmly acknowledges support from the A. von Humboldt Stiftung.

\bibliography{biblio} 
\bibliographystyle{JHEP}

\end{document}